\documentclass[aps,prb,twocolumn,floatfix,superscriptaddress]{revtex4-1}
\usepackage{graphicx,txfonts,bm}
\usepackage[colorlinks,citecolor=magenta,linkcolor=blue]{hyperref}

\begin{document}


\title{Unraveling exotic 5$f$ states and paramagnetic phase of PuSn$_3$}


\author{Haiyan Lu}
\email{hyluphys@163.com}
\affiliation{Science and Technology on Surface Physics and Chemistry Laboratory, P.O. Box 9-35, Jiangyou 621908, China}

\author{Li Huang}
\affiliation{Science and Technology on Surface Physics and Chemistry Laboratory, P.O. Box 9-35, Jiangyou 621908, China}

\date{\today}

\begin{abstract}
Plutonium-based compounds establish an ideal platform for exploring the interplay between long-standing itinerant-localized 5$f$ states and strongly correlated electronic states. In this paper, we exhaustively investigate the correlated 5$f$ electronic states of PuSn$_3$ dependence on temperature by means of a combination of the density functional theory and the embedded dynamical mean-field theory. 
It is found that the spectral weight of narrow 5$f$ band grows significantly and remarkable quasiparticle multiplets appear around the Fermi level at low temperature. A striking $c-f$ hybridization and prominent valence state fluctuations indicate the advent of coherence and itinerancy of 5$f$ states. It is predicted that a 5$f$ localized to itinerant crossover is induced by temperature accompanied by the change in Fermi surface topology. Therefore itinerant 5$f$ states are inclined to take in active chemical bonding, suppressing the formation of local magnetic moment of Pu atoms, which partly elucidates the intrinsic feature of paramagnetic ground state of PuSn$_3$. Furthermore, the 5$f$ electronic correlations are orbital selective manifested themselves in differentiated band renormalizations and electron effective masses. Consequently, the convincing results remain crucial to our understanding of plutonium-based compounds and promote ongoing research.
\end{abstract}


\maketitle

\section{Introduction\label{sec:intro}}

The partially filled 5$f$ electronic shell exhibits itinerant-localized dual nature which remains a long-standing issue in condensed matter physics~\cite{LAReview}. The itinerant 5$f$ states embodied by the noticeable hybridization with conduction bands and the localization degree encoded in magnetic order enable the sensitivity of electronic structure to the external parameters, such as temperature, pressure and chemical doping. Since plutonium (Pu) situates on the edge between the itinerant and localized 5$f$ states of light and heavy actinides~\cite{RevModPhys.81.235}, 5$f$ states tend to involve in active chemical bonding and form affluent Pu-based compounds~\cite{Bauer2015Plutonium}, which exhibit novel quantum phenomena including unconventional superconductivity, magnetic order, nontrivial topology, and heavy-fermion behavior, to name a few~\cite{PhysRevLett.108.017001,PhysRevLett.111.176404,Chudo2013}.
  
\begin{figure}[ht]
\centering
\includegraphics[width=\columnwidth]{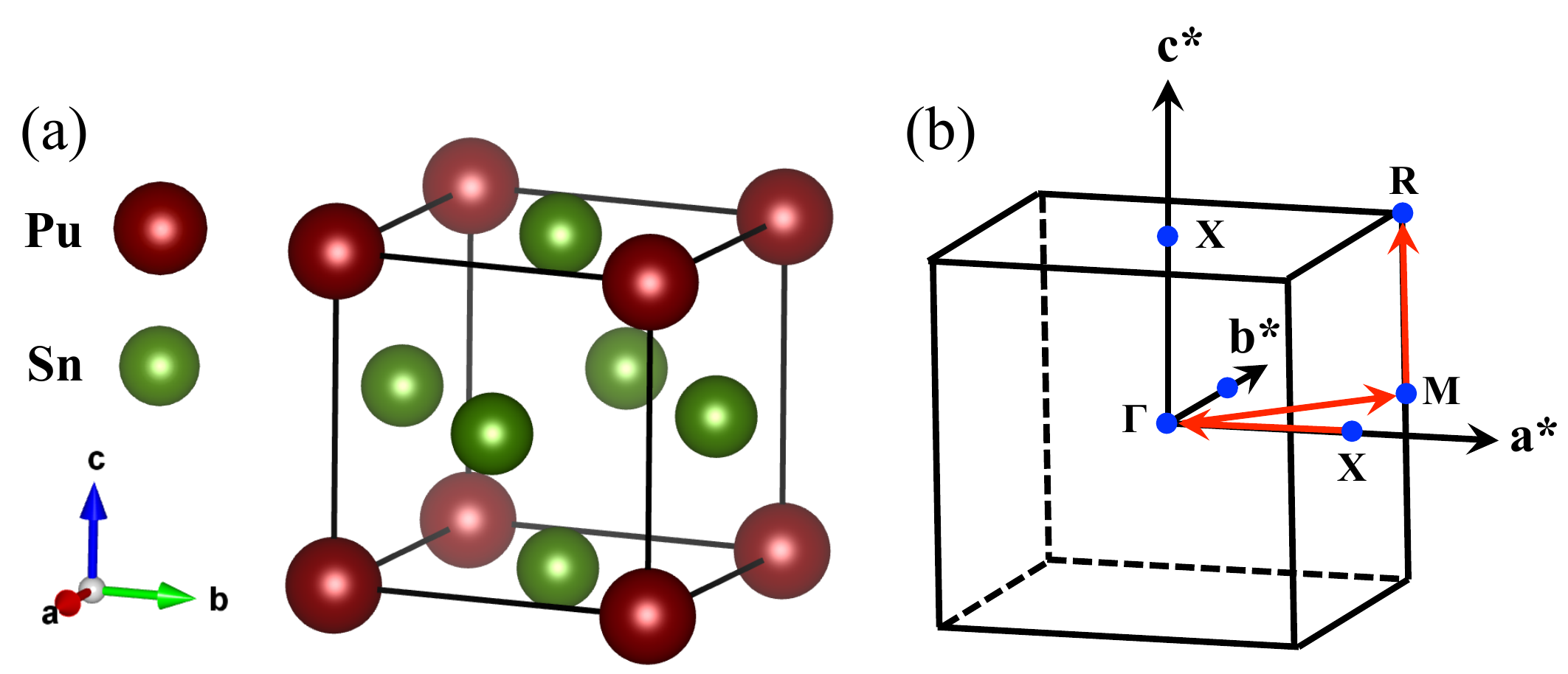}
\caption{(Color online). (a) Crystal structure of PuSn$_3$. (b) Schematic picture of the first Brillouin zone of PuSn$_3$. Some high-symmetry $k$ points $X$ [0.5, 0.0, 0.0], $\Gamma$ [0.0, 0.0, 0.0], $M$ [0.5, 0.5, 0.0], and $R$ [0.5, 0.5, 0.5] are marked.
\label{fig:tstruct}}
\end{figure}

PuSn$_3$ crystallizes in cubic AuCu$_{3}$ structure (space group $Pm$-3$m$) [see Fig.~\ref{fig:tstruct}(a)] with lattice constant 4.63 {\AA}~\cite{SARI1983301}. The temperature-independent paramagnetic order~\cite{Handbookferromag,PhysRevB.39.13115} is pretty anomalous because Pu-Pu distance is much larger than the Hill limit 3.40 {\AA}~\cite{Hilllimit} indicative of localized 5$f$ electrons. In addition, its low-temperature electrical resistivity displays a power-law relation~\cite{Brodsky1978}, which suggests the predominant distribution of conduction $s$ and $p$ states instead of 5$f$ states at the Fermi level. Hence the observed pseudogap at the Fermi level is attributed to the spin-orbit coupling, which enables the fully occupied 5$f_{5/2}$ states and partially distributed 5$f_{7/2}$ states. It is proposed that the $c$-$f$ hybridization becomes the dominant mechanism for Pu-5$f$ electron delocalization~\cite{PhysRevB.39.13115}. So far, photoemission spectroscopy, angle-resolved photoemission spectroscopy, x-ray adsorption spectroscopy and de Haas-van Alphen quantum oscillation experiments are still lacking which altogether provide subtle electronic structure of 5$f$ states.

On the theoretical side, investigations concerning the structural stability, bulk modulus, band structure, electronic density of states and Fermi surface~\cite{PhysRevB.39.13115,BAIZAEE2005247,BAIZAEE2007287,PhysRevB.88.125106} attempt to elaborate narrow 5$f$ bands, Fermi surface topology and pseudogap in density of states at the Fermi level. Even though the Fermi surface and quantum oscillation are exhaustively interpreted, the missing narrow flat 5$f$ bands within the traditional density functional theory is ascribed to the underestimation of strong correlation among 5$f$ electrons~\cite{PhysRevB.88.125106}. Moreover, the unphysical density of states is mostly attributed to the exclusion of the spin-orbit coupling~\cite{BAIZAEE2007287}. Particularly, the underlying mechanism for the paramagnetic ground state has not been clearly understood. More importantly, the strongly correlated electronic states has not been fully quantified. Above all, the temperature-dependent electronic structure is rarely touched in previous literature. Consequently, it is crucial to study the temperature dependence of 5$f$ electronic structure and electronic correlation to disclose the relationship between the electronic structure and paramagnetic ground state of PuSn$_3$.  
 
In the paper, we present the electronic structure of PuSn$_3$ dependence on temperature using the density functional theory in combination with the single-site dynamical mean-field theory. We endeavor to elucidate the itinerant-localized 5$f$ states and to uncover the strongly correlated 5$f$ states. We calculate the momentum-resolved spectral functions, density of states, Fermi surface, self-energy functions, valence state fluctuations and hybridization dynamics of PuSn$_3$. It is found that 5$f$ states become itinerant at low temperature accompanied by remarkable valence state fluctuations. Additionally, the onset of atomic multiplets and prominent $c$-$f$ hybridization at low temperature imply a temperature-driven 5$f$ itinerant-localized crossover. Especially, the change in Fermi surface topology implicates a possible Lifshitz transition induced by temperature. Finally, the orbital selective electronic correlation and hybridization dynamics are addressed.

The rest of this paper is organized as follows. In Sec.~\ref{sec:method}, the computational details are introduced. In Sec.~\ref{sec:results}, the electronic band structures, Fermi surface topology, total and partial 5$f$ density of states, 5$f$ self-energy functions, and probabilities of atomic eigenstates are presented. In Sec.~\ref{sec:dis}, we attempt to clarify some important topics about the 5$f$ itinerant-localized crossover and the hybridization gaps. Finally, Sec.~\ref{sec:summary} serves as a brief conclusion.


\section{Methods\label{sec:method}}
The well-established DFT + DMFT method combines realistic band structure calculation by DFT with non-perturbative many-body treatment of local interaction effects in DMFT~\cite{RevModPhys.68.13,RevModPhys.78.865}. The strong electronic correlation and large spin-orbit coupling are treated on the same footing. 
Here we perform charge fully self-consistent calculations to explore the temperature-dependent electronic structures of PuSn$_3$ using DFT + DMFT method. The self-consistent implementation of this method is divided into DFT and DMFT parts, which are solved separately by using the \texttt{WIEN2K} code~\cite{wien2k} and the \texttt{EDMFTF} package~\cite{PhysRevB.81.195107}. 

In the DFT calculation, the experimental crystal structure of PuSn$_3$~\cite{SARI1983301} was used and ignored the thermal expansion. The generalized gradient approximation was adopted to formulate the exchange-correlation functional~\cite{PhysRevLett.77.3865}. Additionally, the spin-orbit coupling was taken into account in a second-order variational manner. The muffin-tin radius for Pu and Sn were chosen as 2.7 au and 2.4 au, respectively. The $k$-points' mesh was $15 \times 15 \times 15$ and $R_{\text{MT}}K_{\text{MAX}} = 8.0$.  

In the DMFT part, 5$f$ orbitals of plutonium were treated as correlated. The four-fermions' interaction matrix was parameterized using the Coulomb interaction $U = 5.0$~eV and the Hund's exchange $J_H=0.6$~eV via the Slater integrals~\cite{PhysRevB.59.9903}. The fully localized limit scheme was used to calculate the double-counting term for impurity self-energy function~\cite{jpcm:1997}. The vertex-corrected one-crossing approximation (OCA) impurity solver~\cite{PhysRevB.64.115111} was employed to solve the resulting multi-orbital Anderson impurity models. 
 Note that we not only utilized the good quantum numbers $N$ (total occupancy) and $J$ (total angular momentum) to classify the atomic eigenstates, but also made a severe truncation ($N \in$ [3, 7]) for the local Hilbert space~\cite{PhysRevB.75.155113} to reduce the computational burden. The convergence criteria for charge and energy were $10^{-5}$ e and $10^{-5}$ Ry, respectively. It is worth noting that the direct output of OCA impurity solver is real axis self-energy $\Sigma (\omega)$ which was applied to calculate the momentum-resolved spectral functions $A(\mathbf{k},\omega)$ and density of states $A(\omega)$, as well as other physical observables.

\section{Results\label{sec:results}}

\begin{figure*}[th]
\centering
\includegraphics[width=\textwidth]{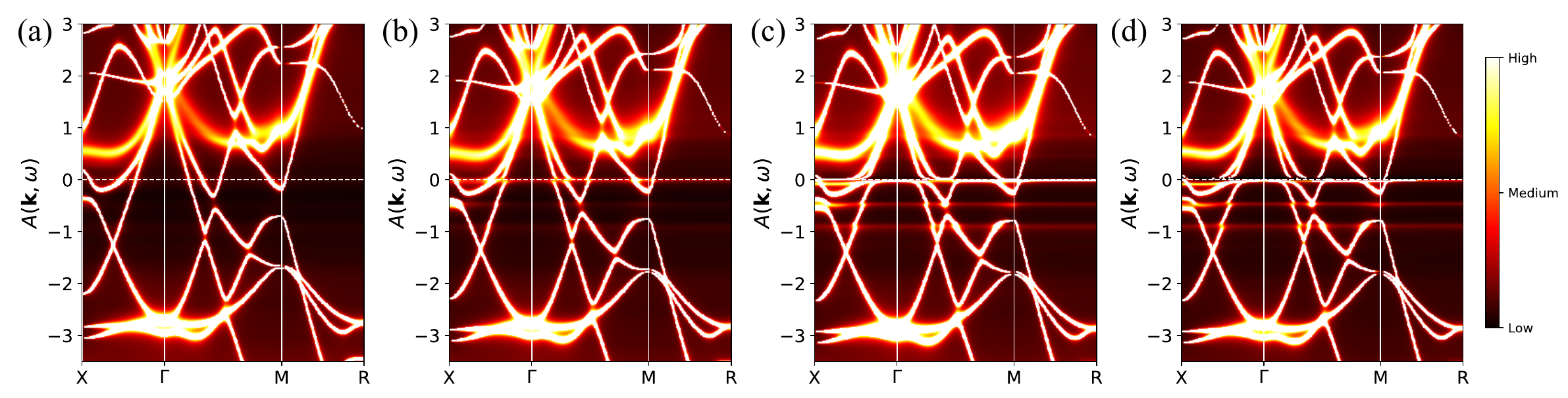}
\caption{(Color online). Momentum-resolved spectral functions $A(\mathbf{k},\omega)$ of PuSn$_3$ as a function of temperature under ambient pressure calculated by the DFT + DMFT method. The horizontal lines denote the Fermi level. (a) 580 K. (b) 290 K. (c) 116 K. (d) 29 K. 
\label{fig:akw}}
\end{figure*}

\begin{figure*}[th]
\centering
\includegraphics[width=\textwidth]{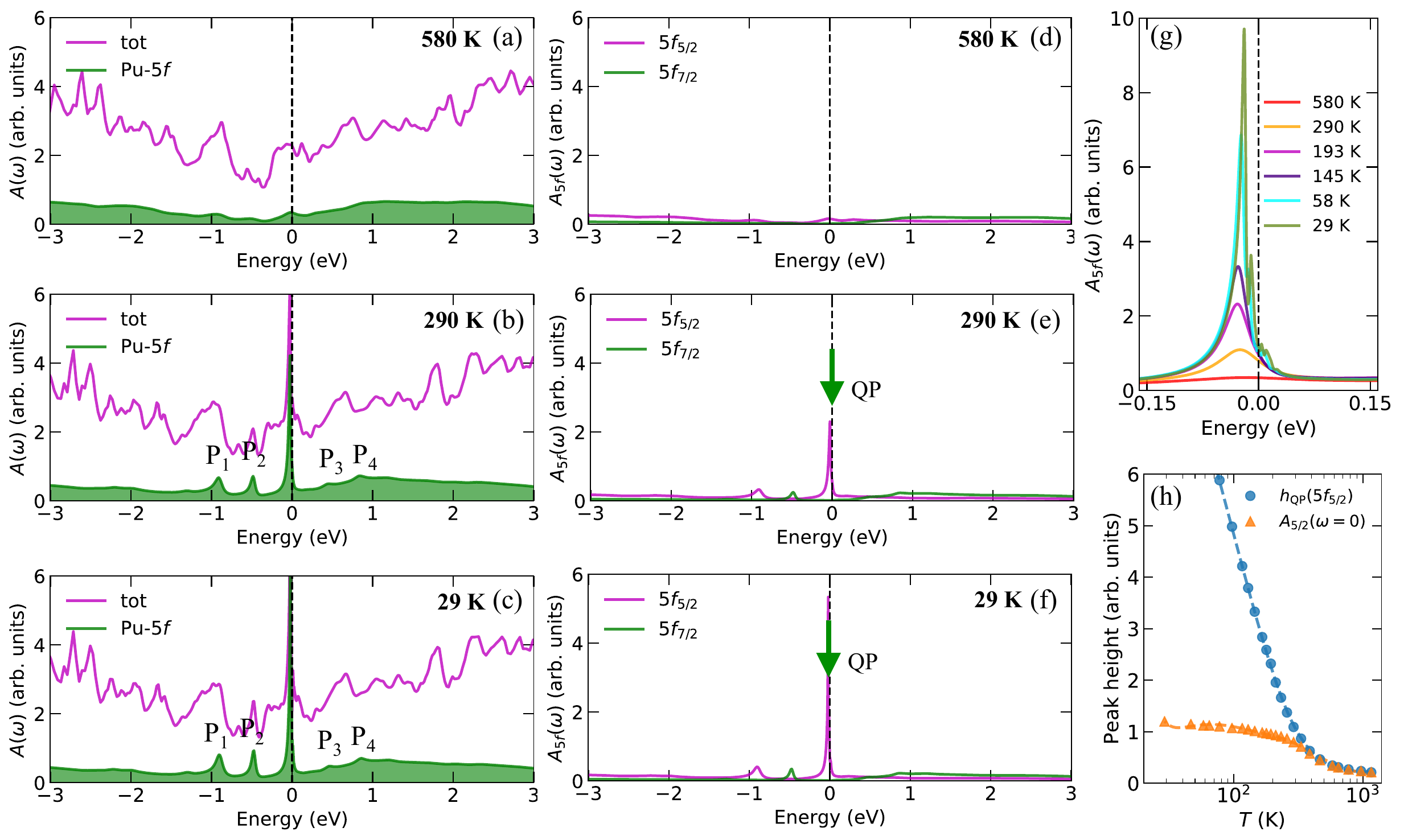}
\caption{(Color online). Electronic density of states of PuSn$_3$ obtained by the DFT + DMFT method.
Total density of states (thick solid lines) and partial 5$f$ density of states (color-filled regions) of 580 K (a), 290 K (b), 29 K (c). These peaks resulting from the quasiparticle multiplets are denoted with ``P1'', ``P2'', ``P3'', and ``P4''. The $j$-resolved 5$f$ partial density of states with $5f_{5/2}$ and $5f_{7/2}$ components represented by purple and green lines, respectively. 580 K (d), 290 K (e), 29 K (f). (g) The evolution of 5$f$ density of states against temperature in the vicinity of Fermi level. (h) The height of the central quasiparticle peak h$_{\rm QP}$(5$f_{5/2}$) and 5$f$ spectral weight at the Fermi level $A_{5f}$($\omega$ = 0) as a function of temperature.
\label{fig:dos}}
\end{figure*}

\begin{figure*}[th]
\centering
\includegraphics[width=\textwidth]{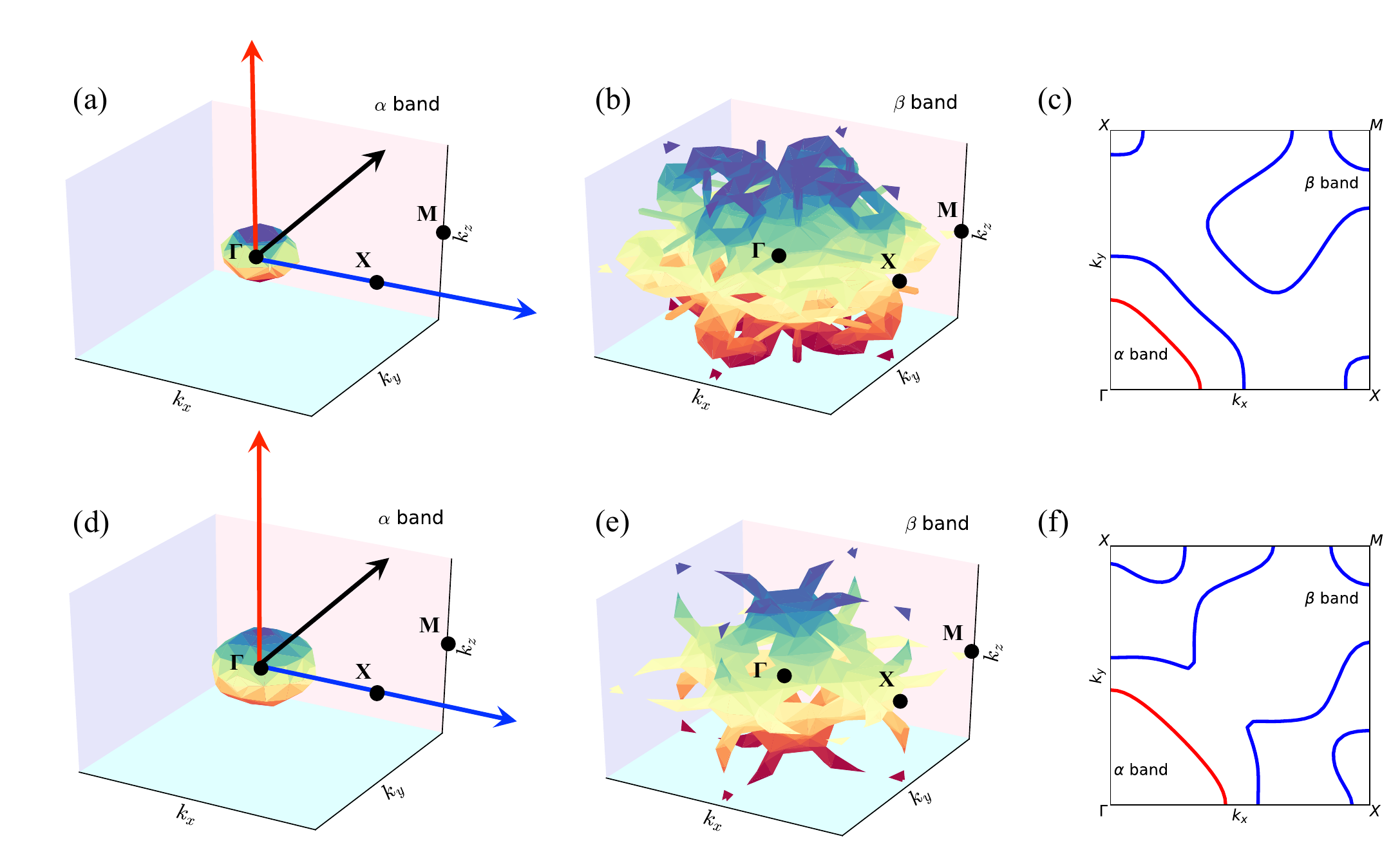}
\caption{(Color online). 
Three-dimensional Fermi surface and two-dimensional Fermi surface of PuSn$_3$ calculated by the DFT + DMFT method at 580 K (a, b, c) and 116 K (d, e, f). Two-dimensional Fermi surface are on the $k_x-k_y$ plane (with $k_z = \pi/2$). They are visualized with different colors.
\label{fig:FS}}
\end{figure*}

\begin{figure*}[th]
\centering
\includegraphics[width=\textwidth]{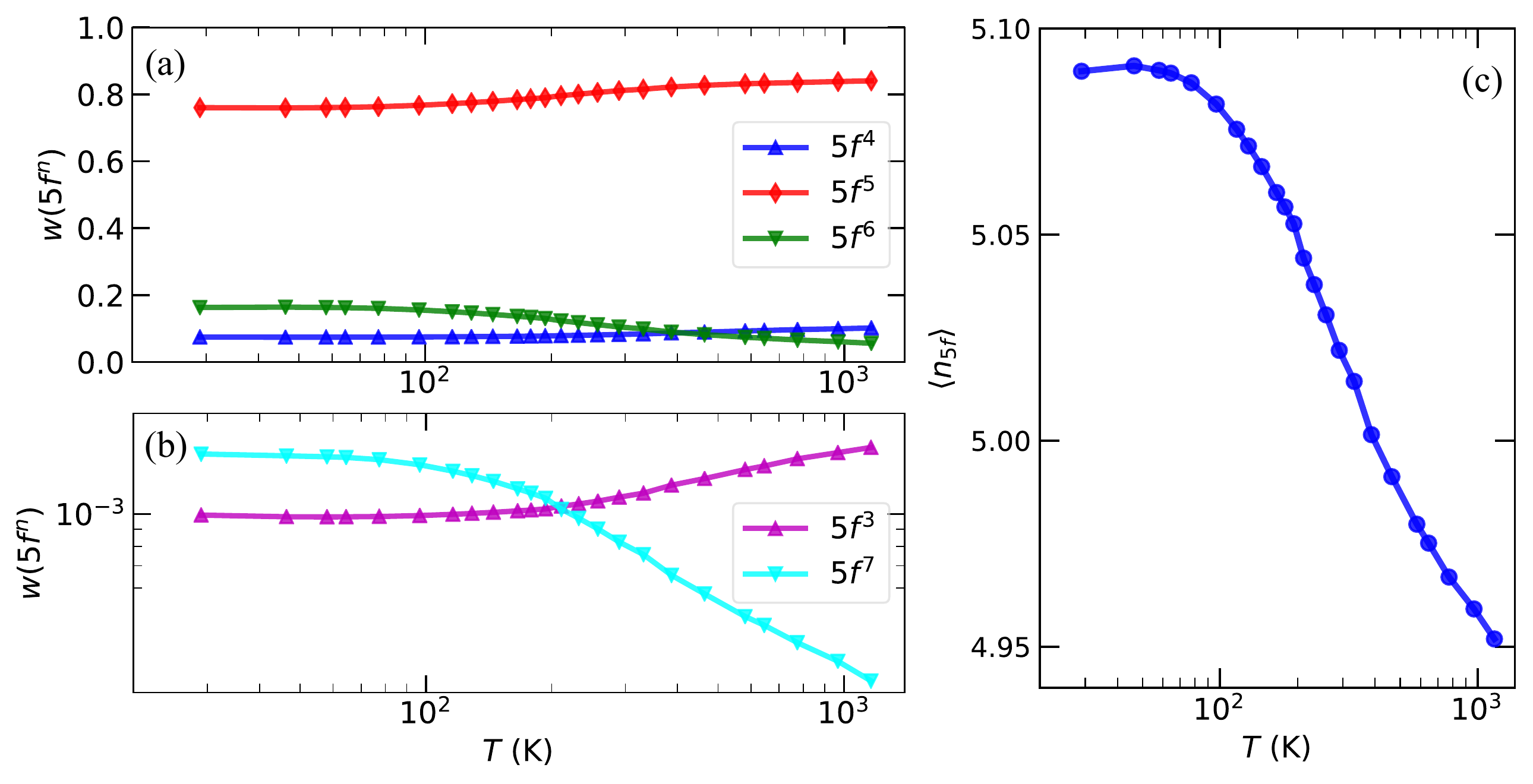}
\caption{(Color online). Probabilities of $5f^{4}$ (blue), $5f^{5}$ (red), $5f^{6}$ (green) configurations (a) and $5f^{3}$ (purple), $5f^{7}$ (cyan) (b) for PuSn$_3$ derived by DFT + DMFT calculations. (c) 5$f$ occupancy as a function of temperature. 
\label{fig:prob}}
\end{figure*}

\subsection{Momentum-resolved spectral functions}
It is illuminating to inspect the momentum-resolved spectral functions which encode interesting features of PuSn$_3$ [see Fig.~\ref{fig:akw}]. We examine the quasiparticle bands as a function of temperature to uncover the inherent nature of 5$f$ states. First of all, it is necessary to evaluate the reliability of our calculated band structures.
In comparison to the band structure of paramagnetic ground state in previous DFT calculations~\cite{PhysRevB.88.125106}, the low temperature momentum-resolved spectral functions [see Fig.~\ref{fig:akw}(d)] share similar characteristics. 
For example, a striking electron pocket locates at $\Gamma$ point about --3 eV with the hole pockets at $X$ and $M$ points below the Fermi level. Generally, the band dispersion around $X$ and $M$ points are reasonably in line with the previous DFT calculations~\cite{PhysRevB.88.125106}. In special, the conduction bands along $\Gamma$ - $M$ high-symmetry line are very much alike. 
Even though the flat bands parallel to the vicinity of Fermi level at low temperature [see Fig.~\ref{fig:akw}(d)] are absent in the DFT calculations, the basic features of energy bands are consistent with each other, evincing the validity of our results. The discrepancy in the existence of narrow 5$f$ bands around the Fermi level indicates the underestimation of strong correlation among 5$f$ electrons within traditional density functional theory, which demands a more rigorous and sophisticated quantum many-body approach. 

Next we turn to the temperature effects on quasiparticle bands of PuSn$_3$. At high temperature [see Fig.~\ref{fig:akw}(a)], only conduction bands with noticeable dispersions cross the Fermi level, accompanied by the emergence of lower and upper Hubbard bands in the energy range of --3 eV $\sim$ --1 eV and 1 eV $\sim$ 3 eV, respectively.
When temperature decreases, 5$f$ electrons gradually get coherent and certain quasiparticle bands start to build up near the Fermi level. These bands are not evident at first because of their small quasiparticle weight and tiny band intensity.
Combined with the density of states in Fig.~\ref{fig:dos}, these flat narrow bands mainly root from 5$f$ states, signifying the itinerant tendency of 5$f$ states. As temperature further lowers, quasiparticle weight grows obvious and apparent narrow 5$f$ bands develop at --0.9 eV, --0.47 eV and the Fermi level, which hybridize with 5$p$ conduction bands of Sn atom to open evident hybridization gaps. It is noted that nearly dispersionless quasiparticle bands are split by spin-orbit coupling with energy gap about 0.43 eV. Meanwhile, the profiles of conduction bands above and below the Fermi level remain almostly unchanged with respect to temperature.  
Accordingly, the significant quasiparticle bands, remarkable $c-f$ hybridization and salient spectral weight of 5$f$ states jointly imply intensifying itinerancy of 5$f$ bands with decreasing temperature. Thus, it is proposed that a potential localized to itinerant crossover of 5$f$ correlated electronic states is induced by temperature in PuSn$_3$.

\subsection{Density of states}
To elucidate the evolution of itinerant degree of 5$f$ correlated electronic states upon temperature, we explore the density of states and quasiparticle multiplets in a wide temperature range of 29 K $\sim$ 580 K. It is noticed that basic features exist in the whole temperature range [see Fig.~\ref{fig:dos}(a)-(c)]. Firstly, the overall profile of total density of states resemble each other including broad ``humps" from --3 eV $\sim$ --1 eV and 1 eV $\sim$ 3 eV, which are mainly assigned to the lower and upper Hubbard bands of plutonium's 5$f$ orbitals. Secondly, the fourteen-fold degenerated 5$f$ states are split into six-fold degenerated 5$f_{5/2}$ and eight-fold degenerated 5$f_{7/2}$ subbands [see Fig.~\ref{fig:dos}(e)-(f)] because of spin-orbit coupling~\cite{RevModPhys.81.235,PhysRevB.102.245111,PhysRevB.101.125123}. A sharp and narrow quasiparticle (QP) peak develops in the Fermi level, mostly belonging to 5$f_{5/2}$ orbital. Meanwhile, the two satellite peaks ``P1'' and ``P2'' at --0.9 eV and --0.47 eV with energy gap about 0.43 eV are ascribed to 5$f_{5/2}$ and 5$f_{7/2}$ orbitals, respectively. 
Above the Fermi level, the reflected peaks ``P3'' and ``P4'' locate at 0.47 eV and 0.9 eV with respect to the central quasiparticle. So the five representative peaks are called ``quasiparticle multiplets''~\cite{2102.02034,2102.03085}. To trace the origin of these quasiparticle multiplets, it is deduced that ``P3'' and ``P4'' peaks are formed by a mix of 5$f_{5/2}$ and 5$f_{7/2}$ orbitals. Overall, it is mentioned that the atomic multiplets are induced by 5$f$ valence fluctuations, which leave fingerprints on the 5$f$ photoemission spectroscopy of PuSn$_3$.  

As to the temperature-dependent coherence and itinerancy of 5$f$ electrons, it is instructive to examine the behavior of quasiparticle multiplets around the Fermi level. It reveals that the quasiparticle peak is too small to be seen in the Fermi level at high temperature, implicating the mostly localized and mainly incoherent 5$f$ states. When temperature slowly descends, the coherence of 5$f$ valence electron gradually builds up. Since the spectral weights of upper and lower Hubbard bands transfer to the Fermi level, the quasiparticle peak around the Fermi level rises up quickly. It is apparent that the height of 5$f_{5/2}$ quasiparticle peak [see Fig.~\ref{fig:dos}(g)] ascends swiftly and acutely, resulting in a sharp and intense quasiparticle peak at the Fermi level. Meanwhile, the quasipartilce weight at the Fermi level progressively magnifies, giving rise to coherent 5$f$ states. In a word, the significant hybridization between 5$f$ electrons of Pu atom and 5$p$ bands of Sn atom suggests the increasing itinerancy of 5$f$ states. The intensifying $c-f$ hybridization contributes to unveil the paramagnetic ground state. Hence the valence state fluctuations become predominant concurrently, which manifests the mixed-valence nature and hints the potential temperature-induced localized to itinerant crossover of 5$f$ states.
 
\subsection{Fermi surface topology}
Fermi surface topology is an effective physical quantity to capture the temperature-dependent electronic structure of PuSn$_3$. Figure~\ref{fig:FS} depicts the Fermi surface topology at two typical temperatures 580 K and 116 K, respectively. There are two doubly degenerated bands crossing the Fermi level (No. of bands: 18 and 19, 20 and 21), which are labeled by $\alpha$ and $\beta$, respectively. Note that the Fermi surface topology of $\alpha$ band takes an ellipsoid shape, which is in coincidence with those in previous DFT calculations~\cite{PhysRevB.88.125106}. Nevertheless, the Fermi surface topology of $\beta$ band resembles an anisotropic form, which acts different from the previous results~\cite{PhysRevB.88.125106}. The discrepancy of $\beta$ band might arise from the temperature effect on Fermi surface because the DFT calculations~\cite{PhysRevB.88.125106} are carried out at zero temperature and we perform the calculations at finite temperature. 
 
As expected, the Fermi surface is sensitive to varying temperature. It is detected that the volume of Fermi surface for $\alpha$ band expands evidently [see Fig.~\ref{fig:FS} (a) and (d)] when temperature diminishes, providing sufficient evidence for the itinerant 5$f$ electrons at low temperature. Since the three-dimensional Fermi surface of $\beta$ band is hard to discern its inner structure, the two-dimensional Fermi surface of $\alpha$ and $\beta$ bands are visualized in Fig.~\ref{fig:FS} (c) and (f). Obviously, both $\alpha$ and $\beta$ bands intersect the $\Gamma$ - $X$ line and the distance between intersections of two bands almost remains unchanged with decreasing temperature, which is in conformity with the momentum-resolved spectral functions [see Fig.~\ref{fig:akw}]. On the contrary, it appears that the distance between the $\beta$ band intersections with Fermi level along the $M$ - $X$ line shrinks as temperature lowers. In particular, the anisotropic Fermi surface topology of $\beta$ band undergoes variation in surface envelope along the $\Gamma$ - $M$ line on the two-dimenional Fermi surface. In consequence, the Fermi surface topology actually changes dramatically with decreasing temperature, which demonstrates a probable Lifshitz transition for 5$f$ states and the advent of 5$f$ localized to itinerant crossover driven by temperature. Since the Fermi surface could be detected by subsequent dHvA quantum oscillation, the experimental results could clarify the underlying mechanism behind the paramagnetic ground state of PuSn$_3$ provided that no signature of Fermi surface nesting is observed. 

\begin{figure*}[th]
\centering
\includegraphics[width=\textwidth]{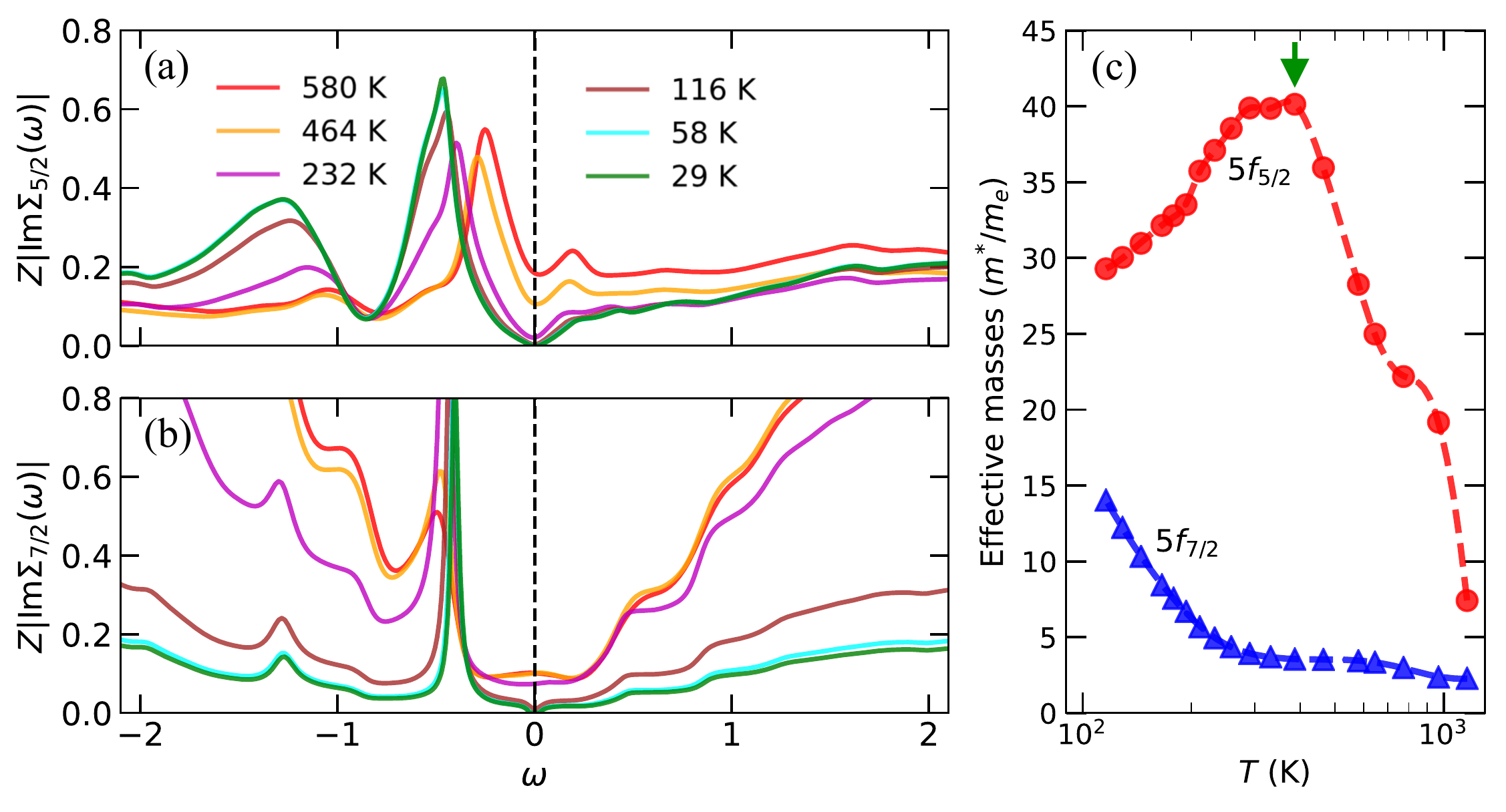}
\caption{(Color online). Real-frequency self-energy functions of PuSn$_3$ obtained by the DFT + DMFT method. (a)-(b) $Z|{\rm Im}\Sigma (\omega)|$ for the 5$f_{5/2}$ and 5$f_{7/2}$ states. $Z$ is the renormalization factor. (c) Electron effective masses for the 5$f_{5/2}$ and 5$f_{7/2}$ states as a function of temperature, where the green arrow denotes maximum electron effective mass for the 5$f_{5/2}$ states.
\label{fig:sig}} 
\end{figure*}

\subsection{Valence state fluctuations}
It is worth mentioning that $\delta$-Pu displays obvious mixed-valence behavior with noninteger occupation number deviating from nominal value 5.0. The 5$f$ electron atomic eigenstates derived from the output of DMFT many-body states shed light on the valence state fluctuations and related mixed-valence behavior.
 Here $p_\Gamma$ is used to quantify the probability of 5$f$ electrons which stay in each atomic eigenstate $\Gamma$. Then the average 5$f$ valence electron is defined as $\langle n_{5f} \rangle = \sum_\Gamma p_\Gamma n_\Gamma$, where $n_\Gamma$ denotes the number of electrons in each atomic eigenstate $\Gamma$. Finally, the probability of 5$f^n$ electronic configuration can be expressed as $\langle w(5f^{n}) \rangle = \sum_\Gamma p_\Gamma \delta (n-n_\Gamma)$. 

The calculated probabilities of 5$f^n$ electronic configuration for PuSn$_3$ are visualized in Fig.~\ref{fig:prob}(a) and (b). Apparently, the probability of 5$f^5$ electronic configuration is dominating, while the contributions of 5$f^3$ and 5$f^7$ electronic configurations are too small to be neglected. As is shown in Table~\ref{tab:prob}, at 580 K, the probability of 5$f^5$ electronic configuration accounts for 83.1\%, followed by the probabilities of 5$f^4$ and 5$f^6$ electronic configurations about 9.3\% and 7.4\%, respectively. Meanwhile, the occupation of 5$f$ electrons is approximately 4.98. When temperature gradually diminishes, the probability of 5$f^5$ electronic configuration declines slowly. At the same time, the probability of 5$f^4$ electronic configuration slightly decreases and the probability of 5$f^6$ electronic configuration appreciably augments. At 29 K, it is noticed that the probability of 5$f^5$ electronic configuration drops to 76.0\%, with 5$f^4$ and 5$f^6$ electronic configurations about 7.4\% and 16.3\%, respectively. At this stage, the occupation of 5$f$ electrons experiences a minor increase to 5.09.
In this scenario, 5$f$ valence electrons are prone to spend more time in the 5$f^6$ configuration, rendering the valence fluctuation and promoting quasiparticle multiplets. In low-temperature regime, it is speculated that PuSn$_3$ is a mixed-valence metal. 
So it is inspiring to refer to isostructual compound PuIn$_3$~\cite{2102.03085}, whose electronic configuration probably suggests insensitivity to varying temperature. Analogous to cubic phase Pu$_3$Ga~\cite{2102.02034}, strong valence fluctuations play a key role in generating the quasiparticle multiplets at the Fermi level, and regulating the effective 5$f$ valence electrons. 

Interestingly, the temperature-dependent patterns of 5$f^4$ and 5$f^6$ electronic configurations intersect about 387 K demonstrating the initiating itinerancy of 5$f$ states and appearance of electron coherence. Coincidently, the electron effective mass for 5$f_{5/2}$ states obtained from self-energy functions according to Eq
.~\ref{eqsigma} reaches the maximum value at 387 K [see Fig.~\ref{fig:sig}(c)]. Generally speaking, electronic correlation is usually strong in localized 5$f$ electrons and grows weaker in itinerant 5$f$ electrons. The coincidence between 5$f$ localized-itinerant nature and strong electronic correlation unravels the hidden connection of itinerant degree and electronic correlation, which requires further exploration.

\begin{table}[th]
\caption{Probabilities of $5f^{3}$, $5f^{4}$, $5f^{5}$, $5f^{6}$, and $5f^{7}$ for PuSn$_3$ at temperatures 580 K and 29 K, respectively.
 \label{tab:prob}}
\begin{ruledtabular}
\begin{tabular}{cccccc}
Temperatures & $5f^{3}$ & $5f^{4}$ & $5f^{5}$ & $5f^{6}$ & $5f^{7}$ \\
\hline
580 K    & 1.357$\times 10^{-3}$ & 0.093  & 0.831 & 0.074 & 4.934$\times 10^{-4}$ \\
29 K   & 9.921$\times 10^{-4}$ & 0.074  & 0.760 & 0.163 & 1.511$\times 10^{-3}$ \\
\end{tabular}
\end{ruledtabular}
\end{table}

\subsection{Self-energy functions}
In general, the electronic correlations are encapsulated in the electron self-energy functions~\cite{RevModPhys.68.13,RevModPhys.78.865}. Figure~\ref{fig:sig} illustrates the renormalized imaginary part of self-energy functions $Z|{\rm Im}\Sigma(\omega)|$ for 5$f_{5/2}$ and 5$f_{7/2}$ states, as well as the electron effective mass.  
Here $Z$ means the quasiparticle weight or renormalization factor, which denotes the electronic correlation strength and can be obtained from the real part of self-energy functions via the following equation~\cite{RevModPhys.68.13}:
\begin{equation}
Z^{-1} = \frac{m^\star}{m_e} = 1 - \frac{\partial \text{Re} \Sigma(\omega)}{\partial \omega} \Big|_{\omega = 0}. \label{eqsigma}
\end{equation}

As is stated above, $Z|{\rm Im}\Sigma(0)|$ embodies the low-energy electron scattering rate~\cite{PhysRevB.99.125113}.
Both $Z|{\rm Im}\Sigma_{5f_{5/2}}(0)|$ and $Z|{\rm Im}\Sigma_{5f_{7/2}}(0)|$ approximate zero at low temperature, suggesting the itinerant 5$f$ states and the metallic feature. And they develop finite values at high temperatues indicating increscent low-energy electron scattering rate. 
With the increment of temperature, $Z|{\rm Im}\Sigma_{5f_{5/2}}(\omega)|$ grows up rapidly and saturates in the energy range of [--0.5 eV, --0.25 eV].
Concurrently, $Z|{\rm Im}\Sigma_{5f_{7/2}}(\omega)|$ climbs up sharply in the high-energy regime ($|\omega|$ $>$ 0.5 eV). Therefore the enhancement of $Z|{\rm Im}\Sigma_{5f_{7/2}}(\omega)|$ becomes more remarkable than that of $Z|{\rm Im}\Sigma_{5f_{5/2}}(\omega)|$, which concretises the growing localization of 5$f_{7/2}$ state and explains its incoherent nature at high temperature. Now that self-energy functions of 5$f_{5/2}$ and 5$f_{7/2}$ states evince differentiated temperature-dependent patterns, it is proposed that 5$f$ electronic correlation are orbital dependent, which commonly exists in actinide compounds.

The evaluated electron effective masses $m^*$ for 5$f_{5/2}$ and 5$f_{7/2}$ states~\cite{RevModPhys.68.13} according to equation~\ref{eqsigma} are given in Fig.~\ref{fig:sig}(c). With the decrement of temperature, the electron effective mass for 5$f_{5/2}$ states surges up rapidly to reach the maximum value up to 40 $m_e$ at 387 K, followed by a monotonically decreasing trend below 387 K. The diminishing electron effective mass for 5$f_{5/2}$ states implies the weakening electronic correlations at low temperature, which is associated with the enhancive itinerancy of 5$f$ states.
Conversely, the electron effective mass for 5$f_{7/2}$ states increases steadily until 387 K. After that it dramatically saturates to a finite value at low temperature. The maximum electron effective mass for 5$f_{7/2}$ states at low temperature hints strong electronic correlation and localized 5$f$ electrons.
The distinct behaviors of electron effective masses for 5$f_{5/2}$ and 5$f_{7/2}$ states are associated with the intrinsic electron correlated characteristic, which confirms the orbital selective correlated states.

\subsection{Discussions\label{sec:dis}}
In this section, we discuss about the heavy-fermion state and temperature-dependent itinerant 5$f$ states to unveil the strongly correlated electronic structure of PuSn$_3$ and underlying mechanism of the paramagnetic ground state. 

\textbf{Electronic heat capacity.}
To explore the heavy fermion behavior of PuSn$_3$, we evaluate the specific heat coefficient within the framework of Fermi-liquid theory. Taking into account the electronic degree of freedom and lattice vibrations, the heat capacity of solid contains two parts $C_v(T)=\gamma T+\beta T^3$. The linear part of heat capacity $\gamma T$ gives the linear specific heat coefficient $\gamma$, which is expressed by the following equation:
\begin{equation}
\gamma = \pi k^2_{B} \sum_{\alpha} \frac{A_{\alpha}(0)}{Z_{\alpha}},
\end{equation}
where $\alpha$ is the orbital index, $A_{\alpha}(0)$ is the spectral weight at the Fermi level, and $Z_{\alpha}$ is the orbital-resolved renormalization factor~\cite{PhysRevLett.101.056403,RevModPhys.68.13}. Since the effective mass enhancement is usually not quite substantial in Pu-based compounds, the specific heat coefficient $\gamma >$ 100 mJ/(mol$\times$K$^2$) is defined as a threshold for Pu-based heavy-fermion compound~\cite{Bauer2015Plutonium}.
Under the neglection of contributions from the conduction bands, the calculated electronic specific heat coefficient of 5$f$ states at 387 K is approximately 96 mJ/(mol$\times$K$^2$). Referring to typical Pu-based heavy-fermion compound PuIn$_3$ with specific heat coefficient 307 mJ/(mol$\times$K$^2$), it is conjectured that PuSn$_3$ might be a promising candidate of Pu-based heavy-fermion compound.

\textbf{Orbital selectivity.}
Here we analyze the temperature-driven itinerant to localized crossover and orbital dependent electronic correlations of 5$f$ states for PuSn$_3$. Primarily, flat narrow quasiparticle bands of 5$f$ states emerge at the Fermi level with amplifying quasiparticle weight and they hybridize with conduction bands to open obvious gaps. Then the itinerancy and coherence of 5$f$ electrons get strengthening at low temperature. 
Due to spin-orbit coupling, 5$f$ states are split into 5$f_{5/2}$ and 5$f_{7/2}$ states, which become itinerant asynchronously. 
Combined with the density of states, it is deduced that the quasiparticle peak rises up quickly at low temperature accompanied by predominant quasiparticle multiplets around the Fermi level. In addition, it is pointed out that quasiparticle multiplets are evoked by valence state fluctuations. It is observed that the central quasiparticle peak is mainly contributed by 5$f_{5/2}$ states which grows coherence at higher temperature than that of 5$f_{7/2}$ states. 
Ultimately, the evolution pattern of 5$f^n$ electronic configuration demonstrates remarkable valence state fluctuations and distinct mixed-valence behavior, which is closely related to the enhancing itinerancy and developing coherence of 5$f$ states. Briefly, it is speculated that temperature-induced localized to itinerant crossover of 5$f$ states is orbital differentiation. 
On the other hand, self-energy functions and electron effective masses of 5$f_{5/2}$ and 5$f_{7/2}$ states display differentiated temperature-dependent behaviors, which signifies the orbital dependent 5$f$ electronic correlation in reminiscence with Pu$_3$Ga~\cite{2102.02034}. 

\begin{figure*}[t!]
\centering
\includegraphics[width=\textwidth]{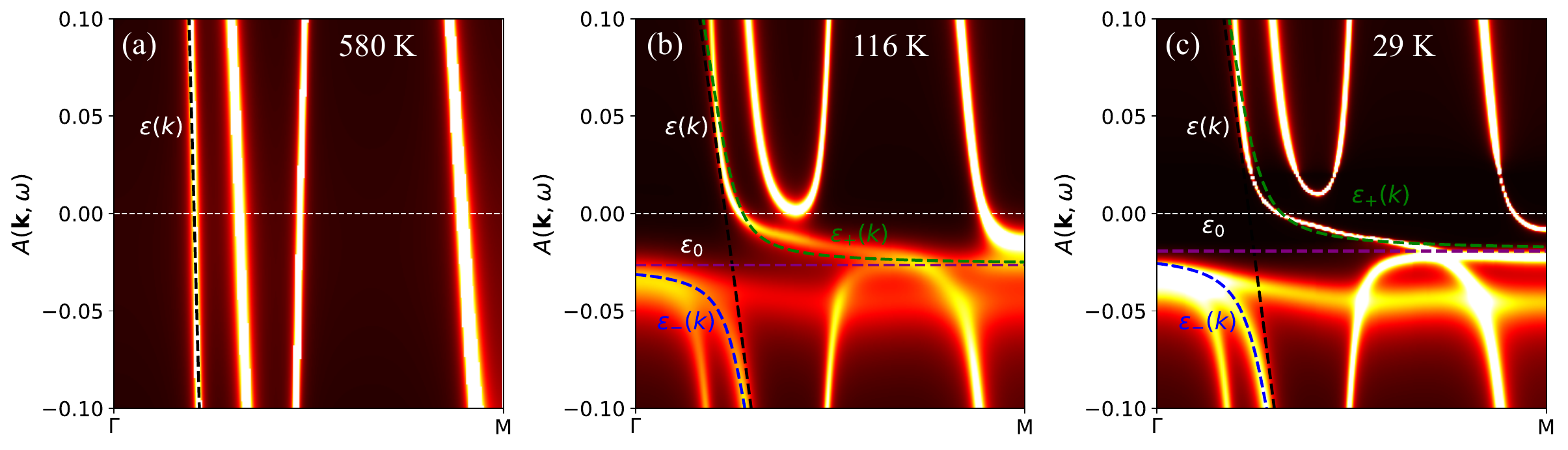}
\caption{(Color online). Temperature dependent quasiparticle band structures (along the $\Gamma$ - $M$ direction) of cubic phase PuSn$_3$ acquired by the DFT + DMFT method. (a) $T$ = 580 K. (b) $T$ = 116 K. (c) $T$ = 29 K.
The white horizontal dashed lines denote the Fermi level. Here, we utilized a periodical Anderson model [see Eq.~\ref{eq:pam}] to fit the low-energy band structures. The colorful dashed lines are the fitting results. The renormalized $5f$ energy level $\epsilon_0$ is denoted by purple line, the unrenormalized band dispersion for conduction electrons $\epsilon(k)$ is plotted in black line. The upper and lower branches of hybridized bands are denoted with green and blue lines, respectively.
\label{fig:hyb}} 
\end{figure*}

\textbf{Evolution of hybridization gap.} 
As is already illustrated in Fig.~\ref{fig:akw}(a)-(d), strong hybridization exists between the 5$f$ bands and conduction bands at low temperature. The $c-f$ hybridization will open a hybridization gap for the conduction bands. Since this gap is adjacent to the Fermi level, the physical properties of PuSn$_{3}$ should be affected largely by it. Hence, it is considerable to determine the size of the hybridization gap and elucidate its temperature dependence relation. Phenomenologically, the low-energy hybridized bands can be well described by a simple mean-field hybridization band picture (i.e. the periodical Anderson model)~\cite{RevModPhys.92.011002}. According to this picture, the energy dispersions are expressed as:
\begin{equation}
\label{eq:pam}
E_{\pm}(k) = \frac{[\epsilon_0 + \epsilon(k)] \pm \sqrt{[\epsilon_0 - \epsilon(k)]^2 + 4|V_k|^2}}{2},
\end{equation}  
where $\epsilon_0$ means the renormalized $5f$ energy level, $\epsilon(k)$ is the unrenormalized band dispersion for conduction electrons, and $V_k$ denotes the strength of hybridization. In the left side of this equation, the ``+'' and ``--'' symbols mean the upper and lower branches of hybridized bands, respectively. In Fig.~\ref{fig:hyb}(a) the band dispersion data at $T = 580$~K are shown. At such a high temperature the hybridization is negligible (i.e. $|V_k| = 0$ and $\epsilon_0 = 0$), so Eq.~(\ref{eq:pam}) is simplified to $E_{\pm}(k) = \epsilon(k)$. Thus, we used the data at $T = 580$~K to calibrate $\epsilon(k)$ (see the black dashed lines in Fig.~\ref{fig:hyb}). The band structures at $T = 116$~K and $29$~K are shown in Fig.~\ref{fig:hyb}(b) and (c), respectively. After fitting Eq.~(\ref{eq:pam}), we derive $\epsilon_0 = -26.5$~meV and $|V_k| = 42.5$~meV for $T = 116$~K, and $\epsilon_0 = -19$~meV and $|V_k| = 50$~meV for $T = 29$~K. Thus, the direct hybridization gaps ($\Delta \approx 2|V_k|$) are 85~meV and 100~meV at 116~K and 29~K, respectively. The results demonstrate that the 5$f$ energy level are pulled toward the Fermi level at low temperature, enhancing the $c-f$ hybridization strength, so as to enlarge the hybridization gap $\Delta$. Similarly, the quasiparticle multiplets also hybridize with the conduction bands. Therefore, multiple hybridization gaps will open at low temperature, which is quite complex and further work is ongoing.


\section{conclusion\label{sec:summary}}
In summary, we studied the detail electronic structure of PuSn$_3$ by employing a state-of-the-art first-principles many-body approach. The temperature dependence of itinerant to localized crossover of 5$f$ states and the correlated electronic states were addressed systematically. As temperature declines, 5$f$ electrons develop coherence and becomes itinerant, exhibiting outstanding quasiparticle multiplets and pronounced valence state fluctuations, accompanied by conspicuous spectal weight and noteworthy $c-f$ hybridization. Especially, the change in Fermi surface topology induced by temperature hints a Lifshitz transition which could be detected by quantum oscillation experiment. Accordingly, 5$f$ states experience a temperature-driven itinerant to localized crossover. Above all, 5$f$ states manifest orbital selective electronic correlation, expressing themselves as orbital dependent electron effective mass and renormalized bands. Our calculated results not only provide a comprehensive picture about how the 5$f$ correlated electronic states evolve with respect to temperature, but also gain deep insight into the complex electronic structure and mysterious paramagnetic state of PuSn$_3$. Further studies about the other Pu-based compounds are undertaken.

\begin{acknowledgments}
This work was supported by the National Natural Science Foundation of China (No.~11874329, No.~11934020, No.~22025602), and the Science Challenge Project of China (No.~TZ2016004). 
\end{acknowledgments}


\bibliography{PuSn3}

\end{document}